\documentclass[aps,prb,twocolumn,superscriptaddress,showpacs]{revtex4-1}

\usepackage{graphicx}
\usepackage{natbib}
\usepackage{bm}
\usepackage{amsmath}
\usepackage{color}
\usepackage[dvipdfm]{hyperref}

\newsavebox{\fmbox}

\begin{document}


\title{Frustrated magnetism in the Heisenberg pyrochlore antiferromagnets $A$Yb$_2X_4$ ($A$ = Cd, Mg, $X$ = S, Se)}



\author{Tomoya Higo}
\email{tomoya@issp.u-tokyo.ac.jp}
\affiliation{Institute for Solid State Physics, University of Tokyo, Kashiwa, Chiba 277-8581, Japan}
\affiliation{CREST, Japan Science and Technology Agency, Kawaguchi, Saitama 332-0012, Japan}

\author{Kensuke Iritani}
\affiliation{Institute for Solid State Physics, University of Tokyo, Kashiwa, Chiba 277-8581, Japan}

\author{Mario Halim}
\affiliation{Institute for Solid State Physics, University of Tokyo, Kashiwa, Chiba 277-8581, Japan}

\author{Wataru Higemoto}
\affiliation{Advanced Science Research Center, Japan Atomic Energy Agency, Tokai, Ibaraki 319-1195, Japan}

\author{Takashi U. Ito}
\affiliation{Advanced Science Research Center, Japan Atomic Energy Agency, Tokai, Ibaraki 319-1195, Japan}

\author{Kentaro Kuga}
\affiliation{RIKEN SPring-8 Center, Sayo, Hyogo 679-5148, Japan}

\author{Kenta Kimura}
\affiliation{Division of Materials Physics, Graduate School of Engineering Science, Osaka University, Toyonaka, Osaka 560-8531, Japan}

\author{Satoru Nakatsuji}
\email{satoru@issp.u-tokyo.ac.jp}
\affiliation{Institute for Solid State Physics, University of Tokyo, Kashiwa, Chiba 277-8581, Japan}
\affiliation{CREST, Japan Science and Technology Agency, Kawaguchi, Saitama 332-0012, Japan}



\begin{abstract}
Our polycrystalline sample study on the Yb-based chalcogenide spinels $A$Yb$_2X_4$ ($A =$ Cd, Mg, $X =$ S, Se) has revealed frustrated magnetism due to the antiferromagnetically coupled Heisenberg spin on the pyrochlore lattice. Our crystal electric field analysis indicates the Yb ground state has nearly Heisenberg spins with  a strong quantum character of the ground doublet.
All the materials exhibit an antiferromagnetic order at 1.4-1.8 K, much lower temperature than the antiferromagnetic exchange coupling scale of $\sim 10$ K.
The magnetic specific heat $C_{\rm M}$ shows a $T^{3}$ dependence, indicating the gapless feature in the Yb-based chalcogenide spinels. The magnetic entropy change much smaller than $R$ ln 2 below the N\'{e}el temperature and the small local magnetic field estimated from the $\mu$SR measurements strongly suggest the significantly reduced size of the ordered moment in comparison with the bare moment size 1.33 $\mu_{\rm B}$/Yb supporting strong fluctuations in the commensurate and incommensurate ordered states in CdYb$_2$S$_4$ and MgYb$_2$S$_4$, respectively.
\end{abstract}


\maketitle



\begin{figure} [t]
\begin{center}
\includegraphics[width=\columnwidth]{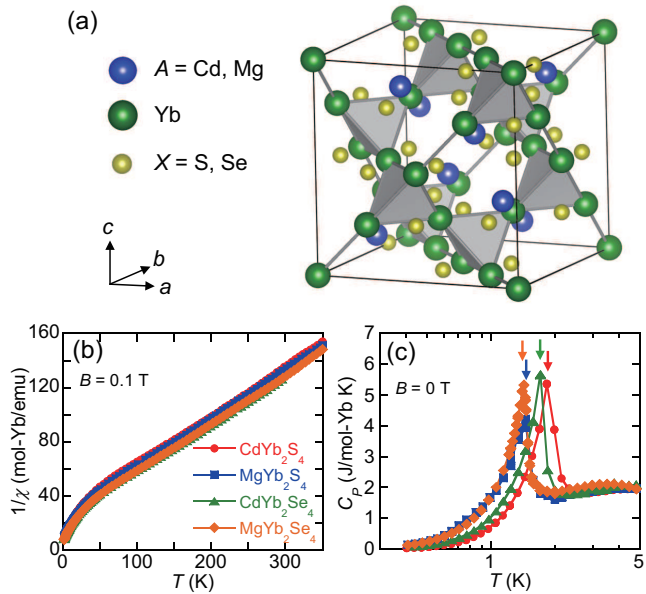}
\caption{(Color online) (a) Crystal structure of the chalcogenide spinels $A$Yb$_2X_4$ ($A$ = Cd, Mg,  $X$ = S, Se). While the tetrahedrally coordinated non-magnetic $A$ sites form a diamond cubic sublattice, the magnetic Yb sites form a pyrochlore lattice with corner-sharing tetrahedra. Temperature dependence of (b) the inverse magnetic susceptibility 1/$\chi$($T$) measured under 0.1 T and (c) the total specific heat $C_{P}$ at 0 T in CdYb$_{2}$S$_{4}$ (red circles), MgYb$_{2}$S$_{4}$ (blue squares), CdYb$_{2}$Se$_{4}$ (green triangles), and MgYb$_{2}$Se$_{4}$ (orange diamonds).} 
\label{CdYbS_crystal}
\end{center}
\end{figure}

Quantum magnetism in geometrically frustrated magnets have attracted great interest in recent years. Various candidate materials of spin liquid states have been discovered in quasi two-dimensional systems based on triangular, Kagome, and honeycomb lattices with a quantum spin $S$ = 1/2. Examples include BEDT-TTF \cite{BEDTTTF}, EtMe$_3$Sb[Pd(dmit)$_2$]$_2$ \cite{dmit}, Herbersmithite \cite{Herbersmithite}, Volborthite \cite{Volborthite}, and Ba$_3$CuSb$_2$O$_9$ \cite{BCSO}.
In three-dimensional systems, the frustrated magnetism on the pyrochlore lattice has been intensively studied \cite{GardnerReview}. One of the most prominent examples is the spin ice \cite{Harris1997,Ramirez1999,bramwell,Castelnovo2012,Gingras2014}, which is based on an Ising spin with a ferromagnetic coupling. Recent theoretical and experimental studies have found that quantum melting of spin ice may lead to a formation of a quantum spin liquid state with emergent topological excitations \cite{Moessner2,Hermele,chiral_spin_liquid,quantum_Tb2Ti2O7,Kim,Machida,Thompson,Ross,Chang2012,Shannon2012PRL,Pr2Zr2O7}.
Spin ice states have been mainly studied using rare-earth based pyrochlore oxides having Ising type $4f$ moments such as Dy$_2$Ti$_2$O$_7$ and Ho$_2$Ti$_2$O$_7$ \cite{Harris1997,Ramirez1999,bramwell,Castelnovo2012,Gingras2014}.On the other hand, various quantum magnetism has been discovered in the pyrochlore oxides having a non-Ising type of ground Kramers doublet. Significant quantum effects have been discussed for the low temperature magnetism found in Yb$_2$Ti$_2$O$_7$ and Er$_2$Ti$_2$O$_7$ \cite{Thompson,Ross,Chang2012,Blote,Champion2003,deReotier,Zhitomirsky,Savary,Ross2014,Petit2014}.
The combination of strong easy-plane anisotropy for low-energy magnetic doublets selected by the local crystalline electric field (CEF) and anisotropic bilinear exchange coupling enhance quantum fluctuations that play an important role in determining the ground state \cite{OnodaPRL,Ross}.

An antiferromagnetic (AF) pyrochlore magnet with isotropic bilinear exchange coupling between nearest neighbor Heisenberg spins has been predicted to host spin disordered ground states in both classical and quantum cases \cite{Moessner,Lacroix1}.
As one of the archetypes of a Heisenberg pyrochlore oxide, Gd$_2$Ti$_2$O$_7$ has attracted much attention and extensive studies have been made and revealed frustrated magnetism with unconventional AF ordering \cite{Raju1999,Champion2001,Ramirez2002,Stewart2004,Brammall2011,Stewart2012}. 

Other than Gd, most of the rare earth ions of pyrochlore oxides are known to have a strong trigonal CEF that stabilizes either Ising or $XY$ planar local symmetry.
To deepen understanding of frustrated magnetism in Heisenberg AF pyrochlore systems, we focus on spinel type compounds $A$$R_{2}$$X_{4}$, where the rare earth $R$ forms the pyrochlore lattice with different coordination from the oxides and possesses a nearly cubic site symmetry that may lead to Heisenberg spins.
While Cd$R_2X_4$ ($X$ = S, Se) systems have been already investigated in the context of geometrical frustration, detailed low temperature measurements have been limited to the $R =$ Er case, where unique magnetic behaviors reflecting the effect of the magnetic anisotropy of rare-earth ions have been found \cite{Lago2010,chalcogenide_spinels,Legros2015,Yaouanc2015}.
On the other hand, nearly cubic site symmetry has been also found in the related Yb-based compound Ba$_3$Yb$_2$Zn$_5$O$_{11}$ \cite{Kimura} with the breathing pyrochlore lattice \cite{breathing_pyrochlore} (an alternating array of small and large tetrahedra). In this compound, the strong quantum effects are manifest in the formation of the spin singlet ground state \cite{Kimura}.

Here, we show that the Yb-based chalcogenide spinels $A$Yb$_2$$X_4$ ($A =$ Cd, Mg, $X =$ S, Se) are good candidate systems of a quantum Heisenberg antiferromagnet on the pyrochlore lattice. To date, only the study at higher temperatures than $\sim$ 2 K has been made on these compounds \cite{Pawlak, chalcogenide_spinels, Pokrzywnicki1974, Pokrzywnicki1975, Ben-Dor1980}. Our low temperature comprehensive experiments have found that the systems exhibit a N\'{e}el order at 1.4-1.8 K, lower temperature than the AF interaction scale of $\sim$ 10 K. Furthermore, our specific heat and $\mu$SR measurements suggest the significantly reduced size of the ordered moment in comparison with the bare moment value due to geometrical frustration and their isotropic character.

\begin{table}[t] 
	\caption{Effective moment $P_{\rm eff}$ estimated from the Curie-Weiss fit for the high-temperature region between 200 and 350 K, Weiss temperature $\theta_{\rm{W}}$ estimated from the Curie-Weiss fit for the low-temperature region between 5 and 20 K, antiferromagnetic transition temperature $T_{\rm{N}}$ estimated from the specific heat, and spin wave velocity $\nu_{\rm{sw}}$ in the chalcogenide spinels $A$Yb$_2$$X_4$ ($A =$ Cd, Mg, $X =$ S, Se).\label{table}}
	\begin{ruledtabular}
		\begin{tabular}{lcccc}
			~~Sample~ &$P_{\rm eff}$ ($\mu_{\rm{B}}$/Yb) &$\theta_{\rm{W}}$ (K) &$T_{\rm{N}}$ (K) &$\nu_{\rm{sw}} \rm{(m/s)}$\\ \hline \hline
			~~CdYb$_{2}$S$_{4}$~ & 4.67 & $-$10.0(4) & 1.8 & 144(2) \\
			~~MgYb$_{2}$S$_{4}$~ & 4.70 & $-$10.4(3) & 1.4 & 100(2) \\
			~~CdYb$_{2}$Se$_{4}$~ & 4.78 & $-$9.3(3) & 1.7 & 131(1) \\
			~~MgYb$_{2}$Se$_{4}$~ & 4.77 & $-$9.2(3) & 1.4 & 107(1)\\
		\end{tabular}
	\end{ruledtabular}
\end{table}

Polycrystalline samples of the Yb-based chalcogenide spinels $A$Yb$_2$$X_4$ ($A =$ Cd, Mg, $X =$ S, Se) are obtained by solid state synthesis \cite{chalcogenide_spinels}. Mixtures of raw elements, Cd(4N) or Mg(3N), Yb(3N), S(5N) or Se(5N) with a ratio 1 : 2 : 4.04, were sintered for a week or two at 900 $^\circ$C twice after sealing them in an evacuated quartz ampule. Our powder X-ray measurements confirmed single phase of the spinel structure for each compound. The lattice constants are obtained as $a = 11.075(1)$ \AA (CdYb$_{2}$S$_{4}$),  $a = 10.972(1)$ \AA (MgYb$_{2}$S$_{4}$),  $a = 11.539(1)$ \AA (CdYb$_{2}$Se$_{4}$),  $a = 11.464(1)$ \AA (MgYb$_{2}$Se$_{4}$), which are consistent with the previous report \cite{chalcogenide_spinels}. The susceptibility was measured using a commercial SQUID magnetometer (MPMS, Quantum Design) between 2 and 300 K under a field of 0.1 T. For the measurements below 2 K down to 30 mK, a home-made SQUID magnetometer in conjunction with a dilution refrigerator was used to measure the DC component under a field of 1 mT, and the AC component with a frequency of 23, 337, 3337 Hz under an AC field of 2.2 $\mu$T. A commercial system (PPMS, Quantum Design) was used to obtain the temperature dependence of the specific heat between 0.4 and 20 K using a relaxation method. Muon spin rotation/relaxation ($\mu$SR) experiments were carried out at J-PARC Muon facility, Tokai, Japan. We have implanted spin polarized positive muons into polycrystalline specimens of CdYb$_{2}$S$_{4}$ and MgYb$_{2}$S$_{4}$,  which should have a similar grain size distribution. The samples were glued on the silver backing holder of the dilution refrigerator by using Apiezon-N grease.


\begin{figure}[t]
\begin{center}
\includegraphics[width=\columnwidth]{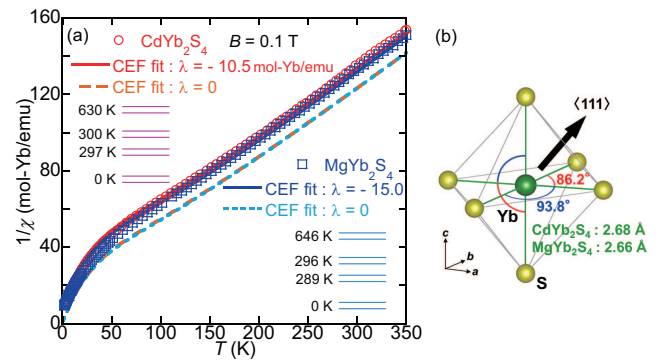}
\caption{(Color online) (a) Temperature dependence of the inverse magnetic susceptibility 1/$\chi$($T$) measured under 0.1 T in CdYb$_{2}$S$_{4}$ (red circles) and MgYb$_{2}$S$_{4}$ (blue squares), compared with  the calculated susceptibility using Eq. (\ref{CEF}) (CEF fit). (Inset) CEF energy level scheme calculated from Eq. (\ref{CEF}). (b) Local environment of Yb$^{3+}$. Crystalline electric field made by six surrounding S$^{2-}$ ions has the symmetry close to cubic with a small trigonal distortion along the $<$111$>$ direction.}
\label{susceptibility}
\end{center}
\end{figure}

\begin{figure}[t]
\begin{center}
\includegraphics[width=\columnwidth]{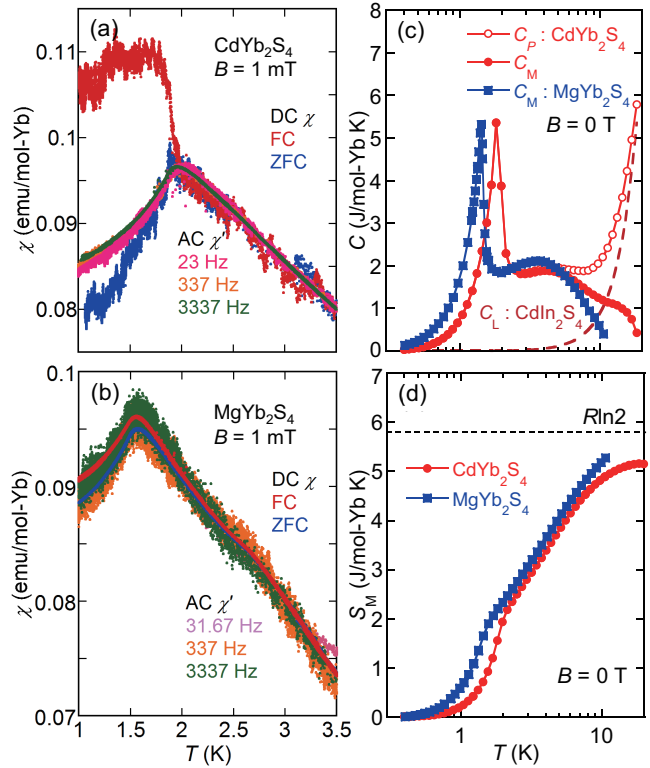}
\caption{(Color online) Temperature dependence of the DC and AC magnetic susceptibility $\chi$($T$) of (a) CdYb$_2$S$_4$, and (b) MgYb$_2$S$_4$. Temperature dependence of (c) the magnetic specific heat $C_{\rm M}$, and (d) the magnetic entropy $S_{M}$ at 0 T of CdYb$_{2}$S$_{4}$ (red circle) and MgYb$_{2}$S$_{4}$ (blue square). The horizontal line indicates the theoretical total entropy for a doublet, $S_{M}$ = $R$ ln 2.}
\label{CTST}
\end{center}
\end{figure}

Let us first discuss the exchange pathways in the chalcogenide spinels.
As shown in Fig. \ref{CdYbS_crystal}(a), the $A$Yb$_2$$X_4$ families have the spinel structure with the space group $Fd\bar{3}m$. The Yb$^{3+}$ (4$f^{13}$) forms the pyrochlore lattice with the sixfold chalcogen $X^{2-}$ coordination and the point symmetry is $D_{3d}$. The nearest neighbor coupling should be based on the superexchange coupling through the Yb-$X$-Yb pathway, as the direct hybridization between Yb $4f$ states is impossible for the nearest neighbor distance (e.g. $\sim 3.9$ \AA{} for  CdYb$_2$S$_4$). 
The second and third neighbor coupling should be significantly reduced in comparison with the nearest neighbor one, as they require a higher order exchange coupling through Yb-$X$-$X$-Yb pathway with a longer distance.
Thus, the spinel chalcogenides can be viewed as model systems to study frustrated magnetism based on the pyrochlore lattice only with an anisotropic nearest neighbor coupling, similarly to the rare-earth pyrochlore oxides.

The chalcogenide spinels $A$Yb$_{2}X_{4}$ exhibit almost the same magnetic behavior.
Figure \ref{CdYbS_crystal}(b) shows the temperature dependence of the inverse susceptibility 1/$\chi$($T$) = $B/M$($T$).
The effective moments $P_{\rm eff}$ obtained from the Curie-Weiss fit for the high-temperature region between 200 and 350 K using the equation, $\chi(T) = C/{(T - \theta_{\mathrm{W}})}$, are closed to the hypothetical value (4.54$\mu_{\rm B}$/Yb) known for an Yb$^{3+}$ ion having a $^{2}$$F$$_{7/2}$ multiplet with $\mathfrak{\textsl{g}}$ = 8/7.
Fittings in the low-temperature region between 5 and 20 K give the negative values of the Weiss temperature $\theta_{\mathrm{W}} \sim -10$ K for all compounds. As we will discuss below, the estimated Weiss temperature is consistent with the exchange coupling $\theta_{\rm W}^{f\!f}$ obtained from the molecular field constant $\lambda$, which indicates the antiferromagnetic coupling between $4f$ magnetic moments. These fitting results are summarized in Table \ref{table}.

To reveal the magnetic ground state, we have measured the specific heat $C_P$ as a function of temperature under zero field. As shown in Fig. \ref{CdYbS_crystal}(c), $C_P$ for all samples clearly exhibits peaks at $T_{\rm N}$ as summarized in Table \ref{table}. The transition temperature of each compound is then determined to be the point where the sign changes in the first derivative of the heat capacity curve with respect to temperature. As we will discuss, these are associated with an antiferromagnetic order.

To further characterize the frustrated magnetism, here we focus on the magnetic properties of the sulfides $A$Yb$_{2}$S$_{4}$ with $A =$ Cd and Mg. First, to determine the local anisotropy of the $4f$ moments, we analyzed the crystalline electric field (CEF) scheme by fitting the temperature dependence of the susceptibility, as shown by solid lines in Fig. \ref{susceptibility}. The following CEF Hamiltonian, 
\begin{eqnarray}
\label{CEF}
\mathcal{H}_{\rm CEF} &=& B_2O^0_2-\frac{2}{3}B_4[O^0_4-20\sqrt{2}O^3_4] \nonumber  \\
&+& \frac{16}{9}B_6[O^0_6+\frac{35\sqrt{2}}{4}O^3_6+\frac{77}{8}O^6_6],
\end{eqnarray}
was used, where $B_{\rm n}$ are the CEF parameters, $O_{\rm n}^{\rm m}$ are Stevens operator equivalents and the $<$111$>$ direction is taken as the quantized axis (Fig. 2(b)).
The exchange coupling effects were taken into account by introducing the molecular field approximation using the 
molecular field constant $\lambda$ as a fitting parameter.
The fitting yields the parameters, $B_2 = 0.3$ K, $B_4 =-0.32$ K, $B_6 =0.0015$ K, and $\lambda = -10.5$ mol-Yb/emu for CdYb$_{2}$S$_{4}$, and $B_2 = 0.5$ K, $B_4 =-0.33$ K, $B_6 =0.0012$ K, and $\lambda = -15.0$ mol-Yb/emu for MgYb$_{2}$S$_{4}$, consistent with previous reports \cite{Pawlak}.
As the exchange parameter is sensitive to sample quality, a small difference in $\lambda$ between our and previous results may arise from e.g. the $A$ and S concentrations in the sample. The small absolute value of $B_2$ indicates the nearly cubic local symmetry of the Yb site, as expected from the local cubic coordination of the chalcogenide anion around an Yb ion (Fig. \ref{susceptibility}(b)). 
The scheme consists of four Kramers doublets, with the excitation energies, 297, 300, 630 K (CdYb$_{2}$S$_{4}$) and 289, 296, 646 K (MgYb$_{2}$S$_{4}$). Thus, the low-temperature magnetism based on the ground doublet should appear at much lower temperatures than room temperature.
The wave function for the ground doublet is obtained as the following form,

\begin{eqnarray}
\label{Wave}
\mathcal{\psi} &=& \mp D \mid \pm \frac{5}{2}> - E \mid \mp \frac{1}{2}> \pm F \mid \mp \frac{7}{2}>,
\end{eqnarray}

\noindent $D$ = 0.804 (0.803) and $E$ = 0.513 (0.515), $F$ = 0.301 (0.299) for ${\rm{CdYb}_{2}\rm{S}_{4}}$ (${\rm{MgYb}_{2}\rm{S}_{4}}$), revealing a strong quantum character of the ground doublet due to the important component mixing including $J_{z} = \pm1/2$.
The ground state wave function has an isotropic Heisenberg spin with longitudinal and transverse components of $\sim 1.33$$\mu_{\rm B}$/Yb. This local isotropy is in sharp constrast with the local Ising anisotropy seen in the other rare-earth spinel CdEr$_{2}$Se$_{4}$ \cite{Lago2010}. 
Given that the CEF ground doublet is made of a pseudo spin-1/2 moment with an effective $\mathfrak{\textsl{g}}$-factor $\mathfrak{\textsl{g}}_{\rm eff} =16/7\langle{J_{z}}\rangle = 2.67$, where $\langle{J_{z}}\rangle$ is the expectation value of the $z$ component of the angular momentum obtained from the wave function for the ground doublet, one can estimate the exchange coupling $\theta_{\rm W}^{f\!f}$ is $-7.0$ K and $-10.0$ K by using the relation $\theta_{\rm W}^{f\!f}$ = $N\mathfrak{\textsl{g}}_{\rm eff}^{2}\mu_{\rm B}^{2}(1/2)(3/2)\lambda/3k_{\rm B}$ \cite{Klimczuk2011} ($\lambda = -10.5$ mol-Yb/emu and $-15.0$ mol-Yb/emu) for CdYb$_{2}$S$_{4}$ and MgYb$_{2}$S$_{4}$, respectively. Here, $N$ is Avogadro's constant. This exchange coupling is consistent with the estimate using the Curie-Weiss fit for the low-temperature region 5-20 K.
The AF coupling $\mathcal{J} = 4.47$ K for CdYb$_{2}$S$_{4}$ and 6.67 K for MgYb$_{2}$S$_{4}$ can be estimated using the relation $|\theta_{\rm W}^{f\!f}| = z\mathcal{J}S(S+1)/3k_{\rm B}$. Here, $z = 6$ is the coordination number.

To reveal the ground state magnetism of $A$Yb$_{2}$S$_{4}$ with $A =$ Cd and Mg, the DC and AC susceptibility measurements were performed using a dilution refrigerator. The results near the transition temperatures are shown in Figs. \ref{CTST}(a) and \ref{CTST}(b). Both of them show anomalies at $\sim$ 2 K.
No frequency dependence was found in the temperature dependence of the AC susceptibility, indicating that the anomalies come from an antiferromagnetic order, not from a spin freezing (Fig. \ref{CTST}).  On the other hand, for CdYb$_2$S$_4$, the DC susceptibility shows a tiny bifurcation below the N\'{e}el  point, $T_{\rm N} = 1.8$ K, which may come from a domain formation, as will be discussed below. 

The $4f$ moment contribution to the specific heat $C_P$ is obtained by subtracting the lattice part estimated by using $C_P$ of the isostructural analogue CdIn$_2$S$_4$ \cite{CdIn2S4}. To account for the volume and mass difference, we followed the same conversion procedure using the Debye equation as described in Ref. \cite{nigasNakatsuji}. Figure \ref{CTST}(c) indicates the temperature dependence of the magnetic part of the specific heat $C_{\rm M}$ under zero field. The specific heat $C_{\rm M}$ clearly exhibits a sharp increase below $\sim 2$ K, and a peak at 1.8 K (1.4 K) for CdYb$_2$S$_4$ (MgYb$_2$S$_4$), indicating a bulk character of the antiferromagnetic transition. The separation between $T_{\rm N}$ and $\theta_{\rm W}^{f\!f}$, quantified by $f = |\theta_{\rm W}^{f\!f}| / T_{\rm N} \sim 4$ for CdYb$_2$S$_4$ and 7 for MgYb$_2$S$_4$, clearly shows the effects of magnetic frustration. A broad peak at 5 K and a tail extending up to $\sim$ 10 K seen in the magnetic part $C_{\rm M}$ may well come from a formation of a magnetic short-range order formed below $T \sim |\theta_{\mathrm W}^{f\!f}|$.

\begin{figure}[t]
	\begin{center}
		\includegraphics[width=\columnwidth]{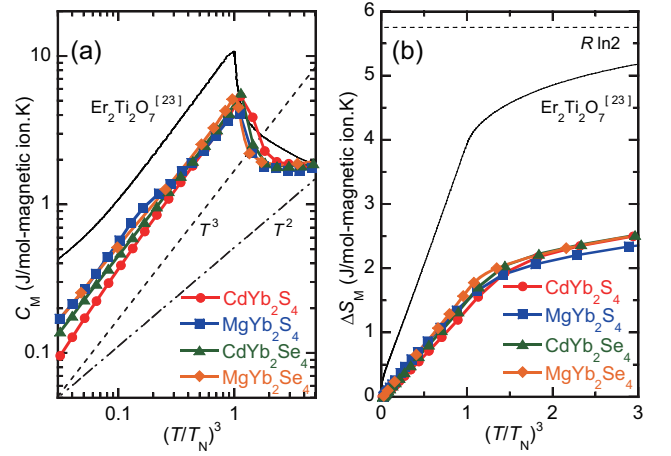}
		\caption{(Color online) (a) Full logarithmic plot of the magnetic specific heat $C_{\rm M}$ vs. $(T/T_{\rm N})^3$ of the chalcogenide spinels compared to Er$_2$Ti$_2$O$_7$ \cite{Blote}, showing nearly $T^3$ dispersion ($C_{\rm M}$ $\propto$ $T^{3.2(1)}$ for CdYb$_{2}$S$_{4}$ (red circles), $T^{2.9(1)}$ for MgYb$_{2}$S$_{4}$ (blue squares),  $T^{3.0(1)}$ for CdYb$_{2}$Se$_{4}$ (green triangles), $T^{3.1(1)}$ for  MgYb$_{2}$Se$_{4}$ (orange diamonds), and $T^{3.1(1)}$ for Er$_2$Ti$_2$O$_7$ in the temperature range $T <$ 0.7 K for MgYb$_{2}$S$_{4}$, $T <$ $T_{\rm N}$ for other spinels, and 0.6 K $\leq T <$ $T_{\rm N}$ for Er$_2$Ti$_2$O$_7$.
			The dashed and two dot dashed lines indicate the slopes for the $T^3$ and $T^2$ laws, respectively. (b) The magnetic entropy $\Delta S_{\rm M}$ vs. $(T/T_{\rm N}) ^3$ of the chalcogenide spinels $A$Yb$_2X_4$ ($A =$ Cd, Mg, $X =$ S, Se) compared to Er$_2$Ti$_2$O$_7$ \cite{Blote}.}
		\label{CmS}
	\end{center}
\end{figure}

Interestingly, in the AF phase below $T_{\rm N}$, the magnetic part of the specific heat $C_{\rm M}$ exhibits the temperature power law behavior, $C_{\rm M}= AT^3$ (Fig. \ref{CmS}(a)).
This gapless feature is expected for a linearly dispersive Nambu-Goldstone mode in three dimensional systems and consistent with the results seen in the electron spin resonance (ESR) measurements for CdYb$_2$S$_4$ \cite{Yoshizawa}.
A similar $T^3$ power law dependence of $C_{\rm M}$ was observed in the antiferromagnetic $XY$ pyrochlore oxide Er$_2$Ti$_2$O$_7$ \cite{Blote, Sosin2010, deReotier}. It should be noted that recent heat capacity and neutron scattering measurements have revealed the presence of a small gap of $\leq$ 0.05 meV ($\sim$0.5 K) in Er$_2$Ti$_2$O$_7$ \cite{deReotier,Ross2014,Petit2014}, while the $T^3$ behavior is observed above the gap energy scale.
Further measurements are required to confirm the presence of the gapless state in the chalcogenide spinels $A$Yb$_2X_4$.

Using the slope $A$, the relation \cite{deReotier}
	\begin{eqnarray}
		A &=& \frac{\pi^2}{120} N_A\frac{k_B^4 a^3}{\hbar^3\nu_{sw}^3},
		\label{SpinWave}
	\end{eqnarray}
allows us to estimate the spin wave velocity $v_{\rm sw}$.
The obtained velocities of the chalcogenide spinels $A$Yb$_2X_4$ are between 100 and 140 m/s (Table \ref{table}).
These values are larger compared to  Er$_2$Ti$_2$O$_7$ (66(1) ms$^{-1}$).
The existence of the linearly dispersive mode in the AF phase put a strong constraint in the determination of the ground state.

Figure \ref{CTST}(d) indicates the temperature dependence of the entropy $\Delta S_{\rm M} = S_{\rm M}(T) - S_{\rm M}({\rm 0.4~K})$ obtained by integrating $C_{\rm M}/T$ vs. $T$ above 0.4 K. The entropy gradually decreases from $R\ln2$ on cooling, consistent with the ground state doublet obtained by the above CEF scheme. The entropy at $T_{\rm N}$ is strongly suppressed due to the formation of short range order at high temperature as a manifestation of the strong geometrical frustration. This entropy suppression is significant when compared to, for example,  Er$_2$Ti$_2$O$_7$ where about $80$\% of $R\ln2$ is released below N\'{e}el temperature \cite{Blote}. Figure \ref{CmS}(b) presents the comparison between both systems in terms of the entropy $\Delta S_{\rm M} = S_{\rm M}(T) - S_{\rm M}({\rm 0.4~K})$. Here we plot the entropy change of the chalcogenide spinels $A$Yb$_2X_4$ and Er$_2$Ti$_2$O$_7$ against $(T/T_{\rm N}) ^3$. All $A$Yb$_2X_4$ compounds show a nearly $T^3$ dependence and the entropy change is suppressed to $\sim 30 \%$ of $R\ln2$ below the N\'{e}el temperature.

\begin{figure}[t]
   \begin{center}
	\includegraphics[width=\columnwidth]{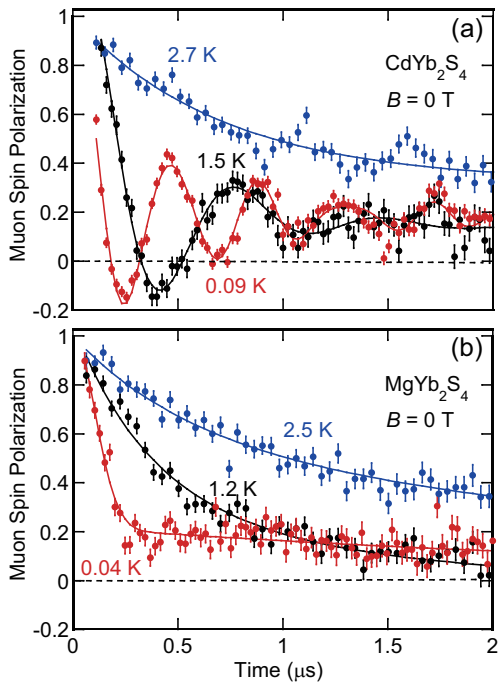}
	\caption{(Color online) $\mu$SR spectra in (a) CdYb$_2$S$_4$, and (b) MgYb$_2$S$_4$ measured at various temperature under 0 T.}
	\label{muSR}
	\end{center}
\end{figure}

In order to further characterize the magnetic ordered state, we performed $\mu$SR measurements, revealing a significantly small local field at the muon site in both CdYb$_{2}$S$_{4}$ and MgYb$_{2}$S$_{4}$ as discussed below.

Figure \ref{muSR}(a) shows the $\mu$SR spectra under zero magnetic field in CdYb$_{2}$S$_{4}$. Spontaneous muon spin precession was seen below 2 K, indicating the static antiferromagnetic ordering state. From the muon spin precession frequency $f$ $\sim$ 2.2 MHz, the local field ( = 2$\pi f / \gamma_{\mu} \colon \gamma_{\mu}$ gyromagnetic ratio of muon) at the muon site is estimated as $\sim$ 15 mT at 0.025 K. Compared CdYb$_{2}$S$_{4}$ with the other $B$-site spinel, the local field is quite small.
For example, in the $B$-site spinel antiferromagnet HgCr$_2$S$_4$\cite{Hige2}, the local magnetic field at the muon site was obtained as $\sim$425 mT. Here, the dipolar field is proportional to the size of the magnetic moment. Moreover, the local magnetic field can be considered to be proportional to the dipolar field. If we assume the same crystal structure, spin structure and muon site, the local field at the muon site should be proportional to the size of magnetic moment. In chalcogenides and oxides, the positive muon is expected to stop at similar sites, near S$^{2-}$ or O$^{2-}$. Therefore, a similar relation between the size of magnetic moments and the local field can be expected in $B$-site antiferromagnetic spinels. A comparison between CdYb$_2$S$_4$ ($\sim$15 mT) and HgCr$_2$S$_4$ ($\sim$425 mT, 3.7 $\mu_{\rm B}$/Cr) suggests the ordered magnetic moment of the order of 0.1 $\mu_{\rm B}$/Yb in CdYb$_2$S$_4$ which is much smaller than the bare moment 1.33 $\mu_{\rm B}$/Yb while it is difficult to estimate the moment size accurately. Similar reduction of the magnetic moment was also suggested in MgCr$_2$O$_4$\cite{Rovers}.

Static magnetic ordering was also observed in MgYb$_{2}$S$_{4}$ below 1.5 K, as shown in Fig. \ref{muSR}(b). However, a spontaneous spin precession was not seen in MgYb$_{2}$S$_{4}$, indicating distribution of the local field. This fact suggests that the magnetic structures in CdYb$_{2}$S$_{4}$ and MgYb$_{2}$S$_{4}$ are different, and MgYb$_{2}$S$_{4}$ has an incommensurate spin order. The $\mu$SR spectra was fitted by using a function
\begin{eqnarray}
P(t) = A_{1}{\rm exp}(-\Delta^{2}t^{2}) + A_{2}{\rm exp}(-\lambda_{\mu} t),
\label{muSR_eq}
\end{eqnarray}
Here $\Delta$ and $\lambda_{\mu}$ indicate the dipolar width at muon site \cite{hayano1979} and dynamic relaxation rate (= $1/T_1$), respectively.  The average local field at the muon site in MgYb$_{2}$S$_{4}$ is obtained as $\Delta$/$\gamma_{\mu} \sim$ 7 mT at 0.04 K.
This value is close to that in CdYb$_{2}$S$_{4}$ and suggests that the ordered moment is much smaller than the bare moment value as is the case in CdYb$_{2}$S$_{4}$.

$\lambda_{\mu}$ shows the critical divergence behavior above $T_N$, indicating a second order phase transition.

\begin{figure}[t]
   \begin{center}
	\includegraphics[width=\columnwidth]{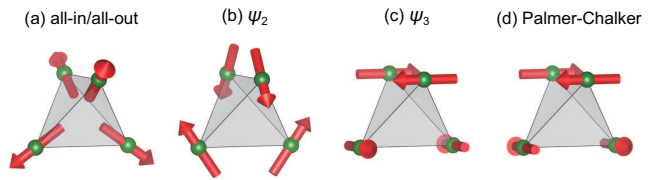}
	\caption{(Color online) Spin arrangements on a tetrahedron in a network of corner-sharing tetrahedra of (a) all-in/all-out, (b) $\psi_2$, (c) $\psi_3$, and (d) Palmer-Chalker state. (a) and (b) are non-coplanar spin antiferromagnetic spin arrengements. (c) and (d) are non-colinear antiferromagnetic spin arrengements.}
	\label{order}
	\end{center}
\end{figure}

While it is hard to identify the spin structure for the incommensurate order suggested for MgYb$_2$S$_4$, there is only a few candidate states that may host a gapless commensurate magnetic order as found for CdYb$_2$S$_4$ based on the combination of the specific heat and $\mu$SR measurements. In these Yb systems, as we discussed above, the superexchange coupling is the dominant intersite magnetic interaction. Given the localized character of the $4f$ moments, the interaction should be dominated by the nearest neighbor coupling.
For a pyrochlore magnet that can be described by anisotropic nearest-neighbor exchange Hamiltonian \cite{OnodaPRL,Ross}, the commensurate order is most likely $q = 0$ type order, which includes
the four types of antiferromagnetic states, as shown in Fig.6, (a) all-in/all-out state ($\Gamma_{3}$); all moments on a tetrahedron in a network of corner-sharing tetrahedra point either inwards or outwards to its center with a non-coplanar spin arrangement, (b) $\psi_2$ state ($\Gamma_{5}$); all the moments are perpendicular to the $\langle111\rangle$ axes and canted symmetrically along the common [100] axis with a non-coplanar spin arrangement, (c) $\psi_3$ state ($\Gamma_{5}$); all the moments are perpendicular to the $\langle111\rangle$ axes and lying in a common (100) plane with a non-collinear spin arrangement, and (d) Palmer-Chalker state ($\Gamma_{7}$); all the moments are perpendicular to the $\langle111\rangle$ axes and arranged in helical manner in a common (100) plane with a non-collinear spin arrangement \cite{yan2017,Poole2007}.
These phases have non-coplanar or non-collinear types of magnetic orders with zero total magnetization on each tetrahedron.
Thus, when the domains are formed below $T_{\rm N}$, the domain wall most likely carries a ferromagnetic component. This may induces a weak ferromagnetic hysteresis between the FC and ZFC susceptibility measurements as observed in CdYb$_2$S$_4$ (Figs. \ref{CTST}(a)).
Experimentally, a similar hysteretic behavior was reported in non-coplanar antiferromagnets which have the pyrochlore and fcc lattices \cite{Petrenko2013,Ishikawa,Matsuhira5,Koda2007,Higo}.

The significantly reduced size of the moment $\ll$ 1$\mu_{\rm_B}$/Yb is striking for a magnetic order in three dimensions. This is consistent with the fact that only 1/3 of $R \ln2$ is released below $T_{\rm N}$, and in sharp contrast to the Er$_2$Ti$_2$O$_7$ case where 0.8$R\ln2$ is released below $T_{\rm N}= 1.25$ K \cite{Blote} (Fig. \ref{CmS}(b)). This indicates that the relatively strong Heisenberg character in comparison with the $XY$ character and the moment size of $3.01\mu_{\rm_B}$ of Er$_2$Ti$_2$O$_7$ \cite{Champion2003} enhances the fluctuation effects and further decrease the ordered moment size while the $XY$ anisotropy in the exchange coupling terms may exist in the chalcogenide spinels $A$Yb$_2X_4$. 

In summary, the rare-earth chalcogenide spinels $A$Yb$_2X_4$ ($A =$ Cd, Mg, $X =$ S, Se) in which Yb forms the pyrochlore lattice with the different coordination from the pyrochlore oxides, has revealed frustrated magnetism due to the antiferromagnetically coupled Heisenberg spin on the pyrochlore lattice. Our CEF analysis indicates the Yb ground state has nearly Heisenberg spins with a strong quantum character of the ground doublet due to the important component mixing including $\pm$$J_{z} = 1/2$. All the materials exhibit antiferromagnetic order at 1.4-1.8 K, much lower temperature than the antiferromagnetic exchange coupling scale of $\sim 10$ K.
The magnetic specific heat $C_{\rm M}$ shows a $T^{3}$ dependence, indicating the gapless feature in the Yb-based chalcogenide spinels. The magnetic entropy change much smaller than $R$ ln 2 below $T_{\rm N}$ and small local magnetic field estimated from the $\mu$SR measurements strongly suggest the significantly reduced size of the ordered moment in comparison with the bare moment value 1.33$\mu_{\rm B}$/Yb supporting strong fluctuations in the commensurate and incommensurate ordered states in CdYb$_2$S$_4$ and MgYb$_2$S$_4$, respectively.

Further measurements such as neutron scattering and electron spin resonance are required to identify the exotic magnetic ground state of the candidate systems of quantum Heisenberg antiferromagnets on the pyrochlore lattice.

\begin{acknowledgments}
We thank D. Yoshizawa, M. Hagiwara, K. Penc, H. Tsunetsugu, Y. Shimura, K.A. Ross, O. Tchernyshyov and C. Broholm for useful discussions. 
We also thank Y. Karaki for low temperature susceptivility measurements. This work is partially supported by CREST, Japan Science and Technology Agency, by Grants-in-Aid for Scientific Research (16H02209) and Program for Advancing Strategic International Networks to Accelerate the Circulation of Talented Researchers (No. R2604) from the Japanese Society for the Promotion of Science (JSPS), and by Grants-in-Aids for Scientific Research on Innovative Areas (15H05882 and 15H05883) from the Ministry of Education, Culture, Sports, Science, and Technology of Japan and by the JSPS Research Fellowship for Young Scientists. 
The use of the facilities of the Materials Design and Characterization Laboratory at the Institute for Solid State Physics, the University of Tokyo, is gratefully acknowledged.
\end{acknowledgments}


%

\end{document}